\documentclass[twocolumn,showpacs,prb,amsmath,amssymb,floatfix]{revtex4}
\usepackage{amsmath,amssymb}
\usepackage{bm}
\usepackage{epsfig}
\usepackage{psfrag}

\newcommand{\bd}{\bm}

\begin{document}

\title{
Two-parameter scaling of  correlation functions
near continuous phase transitions}

\author{Nils Hasselmann, Andreas Sinner, and Peter Kopietz}
  
\affiliation{Institut f\"{u}r Theoretische Physik, Universit\"{a}t
  Frankfurt,  Max-von-Laue Strasse 1, 60438 Frankfurt, Germany}

\date{May 30, 2007}

 \begin{abstract}
We discuss the
  order parameter correlation function
in the vicinity of continuous phase transitions 
using a {\it{two-parameter}} scaling form 
 $G  (  k ) = k_c^{-2}  g ( k \xi , k/ k_c  )$, where $k$ is the wave-vector,
$\xi$ is the correlation length,
and the interaction-dependent non-universal momentum scale $k_c$
remains finite at the critical fixed point. The correlation
function describes the entire critical regime and captures the
classical to critical crossover.
One-parameter scaling is recovered only in the limit $k/k_c\to 0$.
We present an approximate calculation of
 $ g ( x , y )$ for the 
Ising universality class using the functional renormalization group.

\end{abstract}

\pacs{05.70.Fh; 05.70.Jk; 05.10.Cc}

\maketitle

Correlation functions in the vicinity of continuous phase transitions
usually assume a scaling form at long wavelengths.\cite{Ma73,Fisher74} 
Asymptotically, for wave-vectors $k$ much smaller than 
the relevant microscopic scale $\Lambda_0$ 
(such as the inverse lattice
spacing),
the order parameter correlation function $G ( k )$ can be written as
$G ( k ) = k^{-2 + \eta } g^{\pm} ( k \xi )$,
where the anomalous dimension $\eta$ 
characterizes  the universality class of the system,
$\xi$ is the order parameter correlation length, 
and the scaling functions $g^{+} (x )$ and $g^{-} (x)$
describe  the regime above
and below the critical temperature $T_c$, respectively.
Precisely at 
the temperature 
$T = T_c$
there should be a unique scaling function, so that
$g^{+} ( \infty ) = g^{-} ( \infty )$.
If the dimensionality $D$ of the system is smaller than its upper critical dimension, then $g^{\pm} (x )$
approach finite limits for large $\Lambda_0 \xi $ and 
we may take the limit  $\Lambda_0 \rightarrow \infty$.
According to the single-parameter scaling hypothesis, 
the functions $g^{\pm} ( x )$
are essentially determined by their asymptotic limits
for small and large $x$, so that an extrapolation to the
crossover regime $ x\approx 1$ is possible either from
the region $x \gg 1$, or from $x \ll 1$. 
However, it is clear that another prominent scale, 
which is deducible already from a  
dimensional analysis
of Ginzburg-Landau-type theories but is
absent from the one-parameter-scaling picture,
will be important at larger wavevectors: the
interaction-dependent scale $k_c$ which measures the
size of the Ginzburg critical region.\cite{Amit74}
Here we shall assume a microscopic model such that
the scale $k_c$ is small compared to the natural
cutoff of the model. 
This happens e.~g. for Ising
models if the range of interactions is made very large \cite{Kim03}
or near the critical point of complex fluids.\cite{Anisimov05}
In these systems a classical-to-critical crossover can
be observed from a region dominated by the Gaussian
fixed point to the region governed by the Wilson-Fisher
fixed point. While the crossover of static thermodynamic derivates
have been well studied in the past,\cite{Anisimov95,Binder00,Kim03,Anisimov05}
the crossover behavior of the order parameter correlation function
has only been investigated within a one-parameter theory valid right at 
$T_c$.\cite{Baym99,Ledowski04,BlaizotI05}

In this work we extend the one-parameter scaling
theory for correlation functions 
to account also
for 
the interaction-dependent scale $k_c \ll \Lambda_0$ which
divides  the critical regime $ \xi^{-1} \ll k \ll \Lambda_0$
into two separate regimes where  $G ( k )$ has rather different 
properties:\cite{Baym99,Ledowski04,BlaizotI05}
only in the critical long-wavelength (CL) regime $ \xi^{-1} \ll k \ll k_c \ll \Lambda_0$
does the correlation function 
scale asymptotically with an anomalous dimension $\eta$.
For $ k_c \xi \gg 1$ there exists another critical short-wavelength (CS) 
regime $ \xi^{-1} \ll  k_c \ll k \ll  \Lambda_0$
where the behavior of  $G ( k )$ is still universal but  
distinct from the anomalous scaling in the CL regime. 
The scale $k_c$ is present
in the usual renormalization group (RG) analysis, \cite{Anisimov95}
but 
is lost if the RG flow is linearized in the vicinity of the critical
fixed point.

\begin{figure}[tb]
  \centering
  \epsfig{file=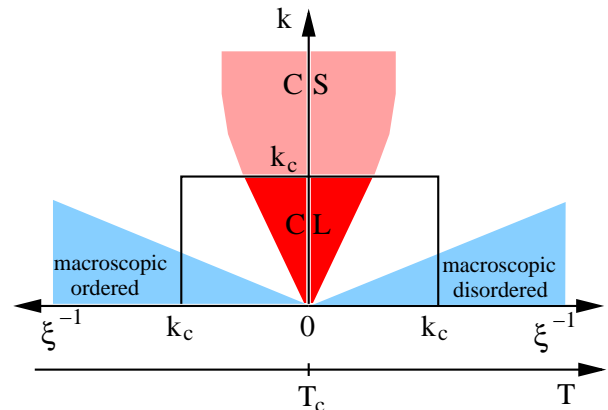,width=80mm}
  \vspace{-4mm}
  \caption{%
(Color online) 
Macroscopic domain of wave-vector $k \ll \Lambda_0$ and correlation length
$\xi^{-1} \ll \Lambda_0$.
In the four shaded regions the correlation function $G (k)$ has
different behavior. The critical regime $ \xi^{-1} \ll k \ll \Lambda_0$
is subdivided into a critical long-wavelength (CL) regime
$\xi^{-1} \ll k \ll k_c $ (where the usual one-parameter scaling
with  anomalous dimension $\eta$ is valid) 
and a critical short-wavelength (CS)  regime $\xi^{-1} \ll k_c \ll  k$
(where the self-energy
exhibits universal power-law scaling distinct from the CL regime).
This figure is a refinement of the corresponding Fig.~1 of
Ref.~[\onlinecite{Halperin68}].
}
  \label{fig:crossover}
\end{figure}

Taking into account the macroscopically ordered and 
disordered regimes,  
four different macroscopic domains of wave-vectors should be distinguished,
as summarized in Fig.~\ref{fig:crossover}.
In order to bring
out the difference between the CL and the CS regime,
we  express  $G (k )$ in terms
of the irreducible 
self-energy $\Sigma ( k )=G^{-1} ( k ) -{k}^2 $.
In the CL regime $\Sigma ( k )  $ scales asymptotically
as $k^{2 - \eta}$ which dominates the 
bare $k^2$-dispersion,
so that $G ( k ) \approx [  \Sigma ( k ) ]^{-1}$.
On the other hand, in the CS regime 
$\Sigma (k)$ may or may not be larger than $k^2$, depending
on the strength of the interaction.
What is more important here is 
that for $k_c \ll k \ll \Lambda_0$
the self-energy can still be expressed in terms of a universal scaling
function which shows power law behavior different from the CL regime.
Therefore,
the self-energy in the macroscopic domain $k \ll \Lambda_0$
should be written in terms  of scaling functions
$\sigma^{\pm} (x,y)$ depending on
{\it{two}} parameters $x = k \xi$ and $y= k/k_c$,
 \begin{equation}
 \Sigma( k ) = k_c^{2} \sigma^{\pm}  ( k \xi , k / k_c )
 \; ,
 \label{eq:scaling}
 \end{equation}
where $\sigma^{+} ( x ,y)$ corresponds to the disordered
phase $T > T_c$,  while $\sigma^{-} ( x ,y )$ describes 
the ordered phase $T < T_c$.
If the system is cooled through $T_c$
and one probes the system at a fixed scale $k$,
then the asymptotic critical behavior
with anomalous exponent $\eta$
can only be observed in  the CL regime $ \xi^{-1} \ll k \ll k_c $.
Yet, the same experiment at
 $ k_c \ll k $ would reveal the universal  behavior of the 
self-energy in the CS regime.
The non-universal scale $k_c$, defining the width of the CL regime,
is missed within the field-theoretical renormalization group, 
which effectively sets $k_c = \infty$ in the self-energy.
However,
$k_c$ and the scaling functions  $\sigma^{\pm}  ( x , y )$
can be calculated using  the functional 
renormalization 
group.\cite{Wetterich93,Morris94,Berges02,Ledowski04,Schuetz06}

\begin{figure}[tb]
  \centering
  \epsfig{file=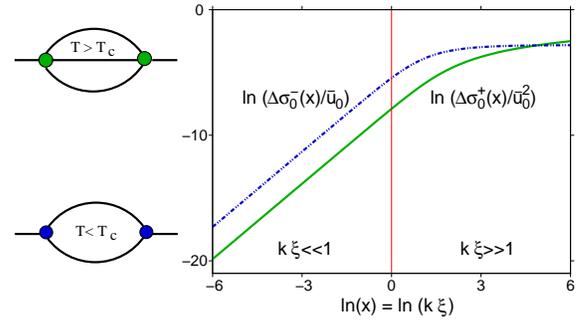,width=75mm}
  \vspace{-4mm}
  \caption{%
(Color online) 
Lowest order perturbative results for the scaling functions $\Delta\sigma_0^{+} ( x)$ (solid line) and
$\Delta\sigma_0^{-} ( x)$ (dashed line) in $D=3$, 
see Eqs.~(\ref{eq:sigmaplus}) and (\ref{eq:sigmaminus}).
The function $\Delta\sigma_0^{+}$ ($\Delta\sigma_0^{-}$)
is determined by the upper (lower)
diagram on the left.
}
  \label{fig:sigmax}
\end{figure}

We believe that the statements above are
general and apply to any
continuous phase transition.
In the rest of this work we 
shall demonstrate their validity
by an approximate calculation of
the scaling functions $\sigma^{\pm} (x , y)$ 
for the Ising universality class in $D$ dimensions,
which can be modeled by an action
involving a real field $\varphi(\bd{r})$,
 \begin{equation}
 S [ \varphi ]  =  
\int d^D{r} \left[ 
\frac{1}{2}({\boldsymbol \nabla} \varphi)^2
+\frac{r_{\Lambda_0}}{2}\varphi^2 
+\frac{u_{\Lambda_0}}{4!}\varphi^4
\right]
,
 \label{eq:Sdef}
\end{equation}
where  
an ultraviolet (UV) cutoff $\Lambda_0$ is assumed to regularize the theory.
It is instructive to consider first the   perturbative
calculation of the self-energy, which is  possible 
as long as the relevant dimensionless coupling  constant
$\bar{u}_0  = u_{\Lambda_0} \xi^{4-D}$ is small.
The correlation length $\xi$ can be expressed in terms
of the self-energy as $ \xi^{-2} = Z \Sigma (0 )$,
where the  field renormalization factor $Z$ is defined as  
$Z^{-1} = 1 + \partial \Sigma ( k ) / \partial k^2 |_{ k =0}$.
For $D < 4$, i.e. below the upper critical dimension of our model, 
the loop integrals
generated in the perturbative expansion are UV convergent and
we may take the limit $\Lambda_0 \rightarrow \infty$.
Perturbation theory 
yields the self-energy in scaling form,
$ \xi^2 \Sigma  (k ) = Z^{-1}+ \Delta\sigma_0^{\pm} ( k \xi )$.
The  lowest order diagrams
giving rise to a momentum dependence of the self-energy
are  shown in Fig.~\ref{fig:sigmax}. 
In the disordered phase the corresponding scaling function is
(up to order $O ( \bar{u}_0^3 )$)
 \begin{equation}
  \Delta\sigma_0^{+} ( x )   =  \frac{ \bar{u}_0^2}{6}
 \int_{\bd{p}}
\chi ( p )
 \left[ \frac{1}{ \bd{p}^2 +1 } -
\frac{1}{ (\bd{p} +   x \bd{n} )^2 +1 }  \right] ,
 \label{eq:sigmaplus}
 \end{equation}
where $\bd{n}$ is an arbitrary unit vector,
 $\int_{\bd{p}} =\int \frac{ d^D p}{ ( 2 \pi )^D}$, and
 $\chi ( p ) =  \int_{ \bd{p}^{\prime}} 
  \big( [\bd{p}^{\prime 2}  +1]  [(\bd{p}^{\prime } + \bd{p} \big)^2 
  +1 ])^{-1}$.
For $T < T_c$
there is a finite three-legged vertex of order $\bar{u}_0^{1/2}$, so that
the leading momentum dependent contribution to the scaling function 
$\Delta\sigma_0^{-} (x)$ is
given by the lower diagram in Fig.~\ref{fig:sigmax} which is linear in 
$\bar{u}_0$,
\begin{eqnarray}
 \Delta\sigma_0^{-} (x)  =
  \frac{ 3\bar{u}_0}{2}  [ \chi ( 0 ) - \chi (x  ) ] + O ( \bar{u}_0^2 ) .
 \label{eq:sigmaminus}
\end{eqnarray}
An explicit evaluation of $\Delta\sigma_0^{\pm} (x)$ in $D=3$ is shown
in Fig.~\ref{fig:sigmax}. 
The qualitative behavior of the  functions $\sigma^{\pm} ( x )$ is 
easily obtained
for arbitrary $D$. While $\Delta\sigma_0^{\pm} (x ) \propto x^2$ for $x \rightarrow 0$,
the asymptote
for large $x$ is 
non-trivial due to the non-analytic behavior
of the function $\chi ( q ) \propto q^{D-4}$ for large $q$ in $D<4$. 
We find $\Delta\sigma_0^{+} ( x ) \propto x^{2 (D-3)}$ for
$3 < D < 4$,  $\Delta\sigma_0^{+} ( x ) \propto \ln x$ for $D=3$, and
 $\Delta\sigma_0^{+} ( x ) \sim \Delta\sigma_0^+ ( \infty ) = O (1)  $ for $2 < D < 3$.
In the ordered phase, the scaling function 
$\Delta\sigma_0^{-} (x)$ approaches for large $x$ a finite limit 
$ \Delta\sigma_0^{-} ( \infty) = \frac{3}{2} \bar{u}_0 \chi (0)$ 
but  the sub-leading correction is non-analytic, 
$\Delta\sigma_0^{-} ( x) - \Delta\sigma_0^{-} ( \infty) \propto x^{D-4}$. 

\begin{figure}[tb]
  \centering
\epsfig{file=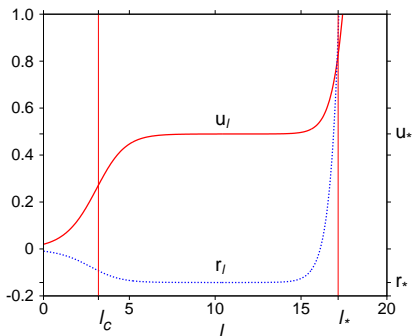,width=62mm}
  \vspace{-4mm}
  \caption{%
(Color online) 
Flow of the coupling parameters ${r}_l$ and ${u}_l$ 
for a nearly critical system 
as a function of the RG time $l$ obtained from 
Eqs.~(\ref{eq:rflow}) and (\ref{eq:uflow}) for $D=3$.
}
  \label{fig:RGflow}
\end{figure}

Lowest order perturbation theory does
not directly reveal the existence of a second characteristic length scale 
$k_c^{-1}$ besides $\xi$. However,
the perturbative approach breaks down 
in the vicinity of the critical point for $D < 4$ since
the effective dimensionless expansion parameter 
$\bar{u}_0 = u_{\Lambda_0} \xi^{4-D}$
diverges. A consistent resummation of these divergences using
a RG
approach not only leads to non-trivial renormalization of $\xi$, but
also reveals the presence of the characteristic scale $k_c$.\cite{Ledowski04}
The signature of both scales $k_c$ and $\xi^{-1}$ is 
already visible in the
usual one-loop RG equations for the rescaled
coupling parameters
${r}_l=Z_l r_\Lambda/\Lambda^2$  and 
${u}_l=K_D Z_l^2 \Lambda^{D-4} u_\Lambda$, describing the evolution of
$r_\Lambda$ and $u_\Lambda$ in Eq.~(\ref{eq:Sdef}) as the  degrees of
freedom in the momentum shell $\Lambda=\Lambda_0 e^{-l} < k < \Lambda_0$
are integrated out.\cite{Ma73,Fisher74} Here, $Z_l$ is a field
renormalization factor and $K_D=2^{1-D}\pi^{-D/2}/\Gamma(D/2)$.
For $T > T_c$ the one-loop RG equations are \cite{Ma73}
 \begin{eqnarray}
 \partial_l {r}_l & = & 2 {r}_l  + 
{u}_l/[ 2 (1 + {r}_l)] \; ,
 \label{eq:rflow}
 \\
 \partial_l {u}_l & = & (4 -D) {u}_l  -  
3 {u}_l^2/[2 (1 + {r}_l)^2].
 \label{eq:uflow}
 \end{eqnarray}
If we fine-tune the initial value ${r}_0$ such that the system
is close to the critical point, the typical RG flow of 
${r}_l$ and ${u}_l$ as a function of the
``RG time'' $l$ is shown in Fig.~\ref{fig:RGflow}.
The two characteristic RG times $l_c$ and $l_{\ast}$ clearly
separate three distinct regimes.
In the regime $  0 < l \lesssim l_c$, the
coupling parameters ${r}_l$ and ${u}_l$ flow toward the values 
${r}_{\ast}$ and ${u}_{\ast}$
associated with the critical fixed point;
in the intermediate interval $  l_c \lesssim l \lesssim l_{\ast} $ the RG 
flow is close to the fixed point and very slow while
for $ l_{\ast} \lesssim l $ the trajectory 
rapidly flows away from the fixed point. 
A simple calculation \cite{Ledowski04} yields the estimate
$l_c \approx ( 4 -D)^{-1} \ln ( {u}_{\ast} / {u}_0)$
for ${u}_0 \ll {u}_{\ast}$.
The one-parameter scaling hypothesis
is based on the analysis of the linearised RG flow in the
vicinity of the fixed point and correlates with the existence of
only one unstable direction. The eigenvalue corresponding
to the unstable direction then directly determines 
the scaling of $\xi$, see e.~g.~Refs.~[\onlinecite{Ma73,Fisher74}]. 
The underlying assumption is that as long as the initial
parameters are close to the critical surface ({\em not} necessarily 
close to the fixed point), 
the unstable direction will completely dominate 
the flow and thus the behavior at small $k$. 
Corrections to pure scaling are known to arise, if the stable (i.e.
irrelavant) directions of the linearised flow are also accounted
for.\cite{Fisher74,Wegner72}
However, the analysis leading to one-parameter
scaling is incomplete 
since the flow
on the critical surface itself will typically introduce (at least)
one additional scale which is not visible in the
linearized flow. 
This is how the characteristic size $k_c$ of the Ginzburg 
critical region enters and how the one-parameter scaling 
hypothesis can be extended.\cite{Anisimov95}
Below, we show that  the RG times $l_c$ and $l_{\ast}$
and the associated scales
$k_c = \Lambda_0 e^{- l_c}$ and $ \xi^{-1} = \Lambda_0 e^{-l_\ast}$
separate regimes where the momentum dependent self-energy 
shows qualitatively different behavior.

\begin{figure}[tb]
  \centering
 \epsfig{file=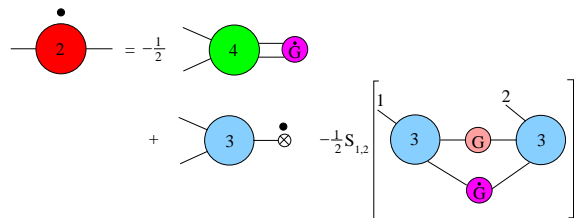,width=75mm}
  \vspace{-4mm}
  \caption{%
(Color online) 
Graphical representation of $\dot{\Gamma}_l ( q )$
in Eq.~(\ref{eq:flowsigma}) describing the effect of higher order
irreducible vertices, represented by numbered circles, 
on  the  evolution of the self-energy 
as degrees of freedom are integrated out.
The crossed  circle represents the flowing  order parameter,
small black dots represent the derivative with respect to $\Lambda$, 
circles with $G$  [$ \dot{G}] $ the exact 
[single scale] propagator, and $S_{1,2}$ symmetrizes
with respect to external labels. For $T>T_c$, the last two 
diagrams involving the
three-legged vertex vanish.
}
  \label{fig:RGdiagram}
\end{figure}

To obtain the  self-energy in the vicinity of the critical point
 taking into account the full RG flow 
 we use the functional renormalization group 
(FRG).\cite{Wetterich93,Morris94,Berges02,Ledowski04,Schuetz06} 
An exact hierarchy of flow equations for the one-line
irreducible vertices of our model is obtained
by differentiating the corresponding generating functional
with respect to $\Lambda$ and then expanding  
this functional in powers of the fields.\cite{Morris94,Schuetz06}
The true self-energy is then obtained as
$\Sigma ( k ) = \lim_{\Lambda \rightarrow 0} \Sigma_{\Lambda} ( k ) $.
To calculate the scaling functions defined in Eq.~(\ref{eq:scaling}), it is
convenient to work with rescaled variables which manifestly exhibit the
scaling dimensions of all 
parameters. 
With 
dimensionless
momenta $ \bd{q} = \bd{k} / \Lambda$ we define the dimensionless self-energy
as $\Gamma_l ( q ) = Z_l \Lambda^{-2} \Sigma_{\Lambda} ( \Lambda q )$, 
where $l = - \ln ( \Lambda / \Lambda_0)$. 
It satisfies an exact flow 
equation of the form \cite{Ledowski04}
\begin{equation}
 \partial_l \Gamma_l ( q )   =  ( 2  -  \eta_l - q \partial_q  ) 
 \Gamma_l ( q )  +  \dot{\Gamma}_l ( q )
 \label{eq:flowsigma}
 \; ,
 \end{equation}
where $ \eta_l = - \partial_l \ln Z_l = \partial \dot{\Gamma}_l ( q ) / \partial q^2 |_{q =0}$
is the flowing anomalous dimension. The
function $\dot{\Gamma}_l ( q )$ depends on higher order
irreducible vertices and
describes the usual mode elimination step of the RG procedure.
A graphical representation of the interaction processes
contained in $\dot{\Gamma}_l ( q )$ 
is shown  in  Fig.~\ref{fig:RGdiagram}.
Note that in the ordered phase, where
our field has a finite vacuum expectation value $\langle \varphi_{\bd{k}} \rangle
 = ( 2 \pi )^D  \delta ( \bd{k} ) M_{\Lambda}$, 
Eq.~(\ref{eq:flowsigma}) has to be augmented by a flow equation
for the flowing order parameter $M_{\Lambda}$,
which is obtained by requiring
that the  vertex with one external leg
vanishes identically for all $\Lambda$.\cite{Schuetz06} 
Defining 
$\dot{\gamma}_l ( q ) =
 \dot{\Gamma}_l ( q ) -  \dot{\Gamma}_l ( 0)$, 
the physical self-energy 
can be written as an integral over the entire RG trajectory,
 \begin{equation}
  \frac{\Sigma ( k )}{\Lambda_0^2}  =  \lim_{l \rightarrow \infty} \frac{ e^{-2l} {r}_l}{Z_l}
+ \int_0^{\infty} d l 
e^{-2 l +\int_0^l d \tau \eta_{\tau}}
 \dot{\gamma}_l ( e^{l} k  /  \Lambda_0 ).
 \label{eq:selfRGres}
 \end{equation}
To relate this to the scaling functions $\sigma^{\pm} ( x , y )$
defined in Eq.~(\ref{eq:scaling}), we use the definitions
$k_c=  \Lambda_0 e^{-l_c}$, $x = k \xi$, $y= k/k_c$. 
By construction
$ e^{ - 2 l_{\ast} } =   ( \Lambda_0 \xi )^{-2} = \lim_{l \rightarrow \infty}  e^{-2 l} {r}_l$ and
 $ x/y = (k_c\xi)=
e^{ l_{\ast}-l_c}$.
Our final result is
 \begin{equation}
 \sigma^{\pm} (x , y )  =  \frac{y^2}{Z x^2} 
+ \int_0^{\infty} d l
 e^{ - 2 ( l - l_c) + \int_0^l d\tau \eta_\tau } \dot{\gamma}_l ( e^{l-l_c} y ).
 \label{eq:selfcrit} 
 \end{equation}
The corresponding
scaling form of the order parameter correlation function
is 
 $G ( k ) = k_c^{-2} g^{\pm} ( k \xi , k /k_c )$, with
 $g^{\pm} (x , y) = [ y^2 + \sigma^{\pm} (x,y) ]^{-1}$.
The behavior of $g^{\pm} ( x , y )$ crucially depends
of the ratio $y/x$. 
Exactly at the critical point one has
$x = \infty$, so that
$\sigma_{\ast} ( y ) = \sigma^{\pm} ( \infty , y )$.
In the CL regime $ y \ll 1$ we obtain the usual anomalous scaling
$\sigma_{\ast} ( y ) \propto y^{2 - \eta}$, while in the CS regime  $y \gg 1$
$\sigma_{\ast} ( y ) \propto \ln y $ in $D=3$, which resembles the
perturbative result $(u_*/\bar{u}_0)^2\Delta\sigma_0^{+} ( k/k_c )$ shown in 
Fig.~\ref{fig:sigmax}
where $k_c$ now plays the role of an effective infrared cutoff.\cite{Baym99}

\begin{figure}[tb]
  \centering
  \epsfig{file=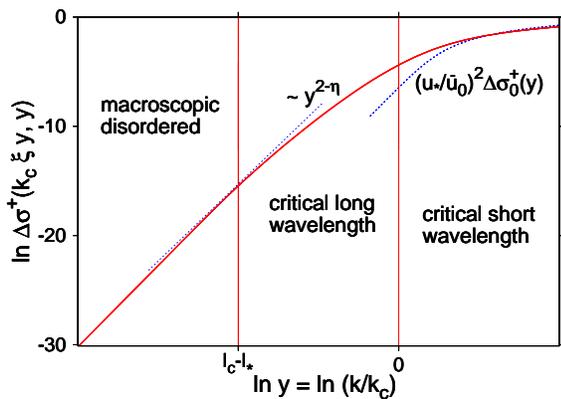,width=75mm}
  \vspace{-4mm}
  \caption{%
Scaling function $\Delta\sigma^{+} ( x  , y )=\sigma^{+} ( x  , y )
-(k_c \xi)^{-2}Z^{-1}$ 
on the line $ x/y = k_c \xi\approx 1700$
in $D=3$ using a truncation of the FRG flow equations
similar to the one used in Ref.~[\onlinecite{Ledowski04}]. Also shown
are $(u_*/\bar{u}_0)^2\Delta \sigma_0^{(+)}(k/k_c)$, which gives a 
reasonable approximation of the CS
regime, and the asymptotic critical behavior
$y^{2-\eta}$ (our truncation leads to $\eta\approx 0.101$)
which is approached within the CL regime before it is replaced by a
$y^2$  behavior in the macroscopic disordered regime. 
}
  \label{fig:selfcrit}
\end{figure}

Away from criticality, Eq.~(\ref{eq:selfcrit}) 
can only be evaluated numerically.
The result of a calculation 
of $\sigma^{+} ( x , y )$  for our model in $D=3$,
using a similar truncation of  the FRG flow equations
as in Ref.~\cite{Ledowski04}, is shown in Fig.~\ref{fig:selfcrit};
the crossover between the different regimes is clearly visible. 
A FRG calculation of $\sigma^{-} ( x, y )$,
using the flow equations for the irreducible vertices in the ordered phase derived
in Ref.~[\onlinecite{Schuetz06}],
will be given elsewhere.\cite{Sinner07}
In the regime $ y \gg x$, which corresponds to $k_c \ll \xi^{-1}$, the system is far away
from criticality (see Fig.~\ref{fig:crossover}), and the scale $k_c$ 
disappears from the problem.
In this case it is better to normalize the self-energy differently,
\begin{equation}
   \xi^2 \Sigma ( k )  = Z^{-1} + \int_0^{\infty} dl e^{ -2 ( l - l_{\ast} ) 
+ \int_0^l d\tau \eta_\tau}
   \dot{\gamma}_l ( e^{l-l_\ast} x ).
 \end{equation}
Evaluating this expression to lowest order in the interaction 
using the trivial scaling ${r}_l \approx   e^{2l} r_{l=0} $ and 
${u}_l  \approx  e^{ (4-D ) l} u_{l=0}$, 
the second term on the right-hand side 
reduces for $ T> T_c$
to the function $\Delta\sigma_0^{+} (x)$ given in 
Eq.~(\ref{eq:sigmaplus}) and for $ T< T_c$ to the function 
$\Delta\sigma_0^{-} (x)$ given in 
Eq.~(\ref{eq:sigmaminus}).

In summary, we argue that the correlation
function  $G(k)$ near continuous phase transitions
has a scaling form which involves (at least) two
parameters.
While in the critical long wavelength (CL) regime the self-energy 
shows asymptotically the usual anomalous scaling, 
$\Sigma ( k ) \propto k^{2 - \eta }$, 
there exists another macroscopic critical short-wavelength (CS)
regime $k_c \ll k \ll \Lambda_0$ 
where the scaling of the self-energy is given by a different power law.
The scaling function describes the crossover between these regimes 
and the macroscopically 
ordered and disordered regime, 
as well as the finite-$k$ corrections to the
asymptotic $k^{2-\eta}$ behavior (which is reached only for $k\to 0$)
in the CL regime. 
The non-universal scale $k_c$ derives from the nonlinear
behavior of the RG flow on the critical surface and is missed
if the flow is linearized around the critical fixed point.
Although for concreteness we have focused here on the Ising
universality class, the two parameter scaling applies to the
critical behavior of 
systems which in the continuum limit are well described
by a Ginzburg-Landau-type action in which terms which are higher 
order than quartic in the fields are negligible. 
The CS regime $k_c \ll k \ll \Lambda_0$ is accessible 
only if $k_c$ is small compared with
the effective UV cutoff $\Lambda_0$. For the case of generalized
Ising-models, this requires that the range of interacions
is very large.\cite{Kim03,Binder00}
A similar crossover might be more readily observed in near-critical
polymer solutions where the crossover scale depends on the radius of
gyration of the polymer \cite{Anisimov05,Binder00} which can be very large.

We thank E.~Vicari and A.~Pelissetto for interesting and useful correspondence.

\end{document}